# Frequency Dependence of the Dielectric Constants and of the Reflectivity for HfO$_2$ and ZrO$_2$ from First-Principles Calculations

C. C. Silva[1], H. W. Leite Alves[1] and L. M. R. Scolfaro[2]

[1]*Departamento de Ciências Naturais, UFSJ, C. P. 110, CEP: 36.300-000 São João del Rei, MG, Brazil*
[2]*Instituto de Física, USP, C.P. 66318, CEP: 05.315-970 São Paulo, SP, Brazil*

**Abstract.** We present, in this work, our theoretical results for the phonon dispersions and the frequency dependence of the reflectivities and the dielectric constants of ZrO$_2$ and HfO$_2$ in the monoclinic phase. The results show the importance of the lattice contribution for the evaluation of the static dielectric constant. Also, besides the anisotropy shown by these materials along the *x* and *z* directions, the zero frequency static dielectric constant decreases with the increasing pressure.

Both Zirconium dioxide (ZrO$_2$) and Hafnium dioxide (HfO$_2$) are good candidates as high-k dielectric materials for gate dielectrics, due to their thermodynamic stability on Si, among other properties. While their structural and electronic properties were extensively studied, the amount of knowledge about their dynamical properties is still rather scarce [1-3]. So, in this work, we have calculated, by using the Density Functional Theory within the Local Density Approximation, plane-wave description of the wavefunctions and the pseudopotential method (abinit code [4]), the phonon dispersions, the hydrostatic strain response of the vibrational modes, and the frequency dependence for both the reflectivities and dielectric constants of both ZrO$_2$ and HfO$_2$ in the monoclinic phase. We have used the Troullier-Martins pseudopotentials, and the phonons were obtained by means of the Density-Functional Perturbation Theory.

Before proceeding with the phonon calculations, we have minimized the total energy of the monoclinic unit cell as a function of the lattice constants and the internal atomic positions. We have then obtained for ZrO$_2$ (HfO$_2$), $a_0$ = 5.12 (5.26) Å, $b_0$ = 5.16 (5.31) Å, $c_0$ = 5.33 (5.44) Å, $\beta$ = 99.6 (99.2) degrees, $B_0$ = 2.44 (2.45) Mbar, and $B_0'$ = 3.9 (4.2), respectively. Our results are in good agreement with the available theoretical and experimental results [3], and they were converged with a cutoff energy of 160 Ry for the plane-wave expansion of the wavefunctions. Also, we have used a (4 4 4) Monkhorst-Pack mesh to sample the Brillouin zone (BZ).

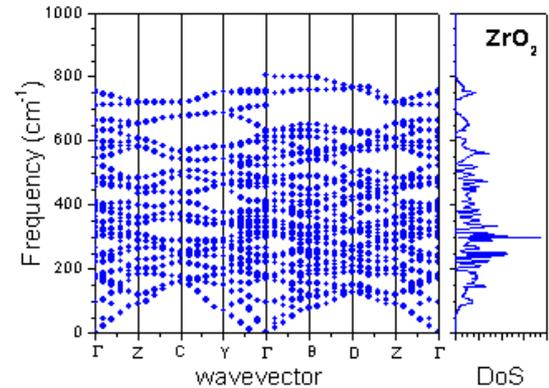

**FIGURE 1.** Phonon dispersions and density of states calculated for the monoclinic ZrO$_2$.

In Figure 1, we display our calculated phonon dispersions as well as the density of states for the ZrO$_2$. Our results agree well with both the Raman data and other theoretical results and, we did not find any mode with negative value for its frequency, as observed for its cubic and tetragonal modifications [5]. However, when applying hydrostatic pressure, five modes have shown a soft mode character, $1A_g$ (at 96 cm$^{-1}$), $1B_u$(TO) (at 219 cm$^{-1}$), $2A_u$(TO) (at 242 cm$^{-1}$), $2A_u$(LO) (at 242.5 cm$^{-1}$) and $3B_g$ (at 308 cm$^{-1}$): they have a negative value for their Grüneisen parameter $\gamma$, which is directly related to the instability of this structure. Besides that, the other modes show $\gamma$ values greater than 1, reflecting a flat behaviour for their dispersions close to the Γ point of the BZ.

The obtained phonon dispersions and density of states for $HfO_2$ are depicted in Figure 2. They show the same features observed for the $ZrO_2$ with frequencies smaller (~ 3%) when compared with the latter.

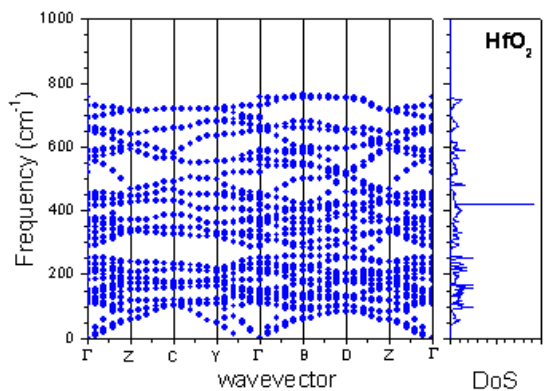

**FIGURE 2.** Phonon dispersions and density of states calculated for the monoclinic $HfO_2$.

In Figure 3, we show the calculated hydrostatic pressure dependence of the components of the static dielectric constant, $Re\{\varepsilon(0)\}$, for the $ZrO_2$ (the results for the $HfO_2$ show the same features). It is interesting to remark that the high average values obtained for the static dielectric tensor (20.78 for $ZrO_2$, and 18.02 for $HfO_2$) are indications of the importance of the lattice contribution for the evaluation of this tensor, once its experimental value is around 25 [6]. From Figure 3, we have also noted that all the components of the $Re\{\varepsilon(0)\}$ decrease with the increasing pressure. We can infer, then, that thin films of these oxides can exhibit lower values for their dielectric constants depending of the interface strains with Si substrate.

Finally, in Figure 4, we show the obtained frequency dependence of the reflectivity for the $ZrO_2$. From Figure 4, the strong anisotropy of the oxide along the x and z directions of the crystal is due to the different responses of the $B_u$ modes with the incoming electromagnetic waves. Only the $B_u$ mode at 220 cm$^{-1}$ is excited in both these directions, as shown by the coincidence of the first peak of both $R_x$ and $R_z$ curves. Moreover, $R_x$ has four Reststrahlen bands, while $R_y$ and $R_z$ have five ones. For the $R_y$ bands, only the $A_u$ modes at 247, 473 and 576 cm$^{-1}$, respectively, are excited (the other two are $B_u$ ones), while the $R_x$ and $R_z$ bands are only formed by some of the $B_u$ modes.

In summary, we have presented our obtained results for some dynamical properties of both $ZrO_2$ and $HfO_2$, such as phonon dispersions, dielectric constants and reflectivities. A complete description of our obtained results will be published soon elsewhere.

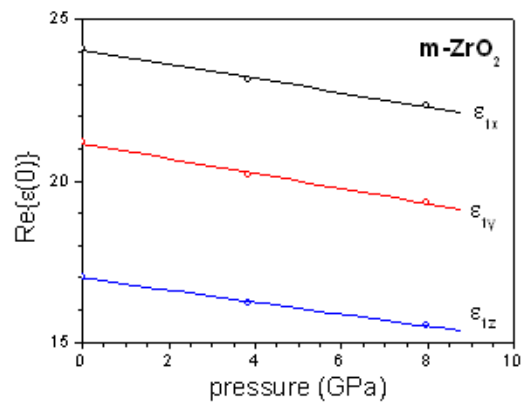

**FIGURE 3.** Pressure dependence of the components of the static dielectric tensor evaluated for the $ZrO_2$.

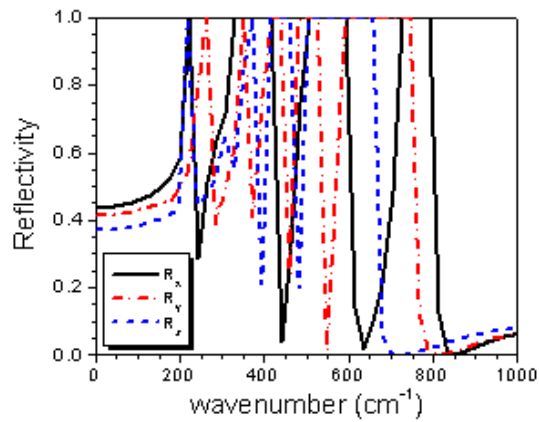

**FIGURE 4.** Frequency dependence of the reflectivity calculated for the $ZrO_2$.

## ACKNOWLEDGMENTS

This work was supported by the FAPEMIG project CEX 80953/04, Brazil.

## REFERENCES

1. S. N. Tkachev *et al.*, *Spectrochim. Acta* **A61**, 2434 (2005), and references therein.
2. G. A. Kouroukiis and E. Liarokapis, *J. Am. Chem. Soc.* **74**, 520 (1991).
3. X. Zhao and D. Vanderbilt, *Phys. Rev.* **B65**, 075105 (2002); *Phys. Rev.* **B65**, 233106 (2002), and references therein.
4. X. Gonze *et al.*, *Comput. Mater. Sci.* **25**, 478 (2002), and references therein.
5. F. Detraux *et al.*, *Phys. Rev. Lett.* **81**, 3297 (1998).
6. J. C. Garcia *et al.*, *J. Appl. Phys.* (2006), accepted.